# Multislice Electron Tomography using 4D-STEM

Juhyeok Lee[1], Moosung Lee[1,2], YongKeun Park[1,2,3], Colin Ophus[4] and Yongsoo Yang[1,*]

[1]*Department of Physics, Korea Advanced Institute of Science and Technology (KAIST), Daejeon 34141, Korea*
[2]*KAIST Institute for Health Science and Technology, Daejeon 34141, Korea*
[3]*Tomocube, Inc., Daejeon 34051, Korea*
[4]*National Center for Electron Microscopy, Molecular Foundry, Lawrence Berkeley National Laboratory, Berkeley, California 94720, United States*

[*]Email: yongsoo.yang@kaist.ac.kr

**ABSTRACT**

Electron tomography offers important three-dimensional (3D) structural information which cannot be observed by two-dimensional imaging. By combining annular dark field scanning transmission electron microscopy (ADF-STEM) with aberration correction, the resolution of electron tomography has reached atomic resolution. However, tomography based on ADF-STEM inherently suffers from several issues, including a high electron dose requirement, poor contrast for light elements, and artifacts from image contrast nonlinearity. Here, we developed a new method called MultiSlice Electron Tomography (MSET) based on 4D-STEM tilt series. Our simulations show that multislice-based 3D reconstruction can effectively reduce undesirable reconstruction artifacts from the nonlinear contrast, allowing precise determination of atomic structures with improved sensitivity for low-Z elements, at considerably low electron dose conditions. We expect that the MSET method can be applied to a wide variety of materials, including radiation-sensitive samples and materials containing light elements whose 3D atomic structures have never been fully elucidated due to electron dose limitations or nonlinear imaging contrast.

## I. INTRODUCTION

Electron tomography is a powerful microscopy technique for three-dimensional (3D) analysis of various materials at the nanoscale spatial resolution, with applications in diverse fields, including physics, chemistry, biology, material sciences, and medicine [1,2]. Quantitative structural characterization in 3D provides direct insights into the morphologies, compositions, and functionality of complex materials. Aided by advances in aberration correction technology and tomographic reconstruction algorithms, the resolution of tomograms has been continuously improved [3,4]. In recent years, atomic resolution tomography has been demonstrated via annular dark field scanning transmission electron microscopy (ADF-STEM) [5–13].

ADF-STEM tomography allows us to measure the 3D atomic positions of various nanomaterials at the single-atom level with elemental specificity; combined with quantum mechanical calculations such as density functional theory (DFT), their structure-property relations can directly be obtained at the most fundamental level. Since electron tomography does not rely on the assumption of crystallinity, it allows atomic-scale determination of 3D non-crystalline structures such as defects, dislocations, and chemical order/disorder, which cannot be easily detected by X-ray crystallography or 2D imaging [5,7,10,14]. However, ADF-STEM imaging requires a relatively high electron dose due to the low detection efficiency of ADF, which counts only high-angle scattered electrons, and therefore radiation damage substantially limits its applicability for radiation-sensitive materials. Additionally, the contrast of ADF images is approximately proportional to $Z^{1.7}$ where Z is the atomic number [15–18], making it difficult to identify light elements, particularly when heavier elements are adjacent to them. In addition, multiple scattering effects can cause nonlinear imaging contrast in ADF-STEM images [19], which can deteriorate the resolution and create undesirable artifacts in the 3D reconstruction, arising from violations of the linear projection rule required by typical tomography reconstruction algorithms.

Alternative approaches have been suggested to avoid these issues. Atomic resolution tomography based on bright-field high-resolution transmission electron microscopy (HRTEM) has been reported for very small crystals in liquid cell environments [20,21]. This method works with relatively lower electron dose without needing to tilt the sample holder. However, the method still suffers from strong nonlinear effects, which harm the reconstruction fidelity for larger samples. Ren and his coworkers have proposed a phase contrast retrieval algorithm to reconstruct the 3D atomic potentials from HRTEM images taken at multiple tilt angles and defocus values [22,23]. This approach considers multiple scattering of an electron beam and phase shifts induced by scatters, which is related to the algorithms considering multiple scattering of light in 3D quantitative phase imaging [24–28]. Their simulations showed that atomic resolution 3D reconstruction of atomic potentials could be obtained from nonlinear HRTEM images under low electron dose conditions. However, it requires accurate registration of focal stacks, which is challenging to achieve for low-dose experimental datasets.

Ptychography-based electron tomography using 4D-STEM (referring to 2D diffraction images of a converged electron beam at each point in a 2D STEM raster [29,30], see Fig. 1a) has also been recently demonstrated at the atomic scale, which takes advantage of the linear contrast that can be obtained from ptychographic reconstruction of 4D-STEM measurements [31,32]. In this approach, 4D-STEM-based ptychography data is acquired for each tilt angle. 2D ptychographic images are then reconstructed via ptychographic algorithms such as single sideband (SSB) or extended ptychographic iterative engine (ePIE) [33,34]. The tomographic reconstruction algorithm ensues to convert the reconstructed 2D images to generate 3D tomograms. Since ptychography image contrast scales linearly with the atomic number, it can facilitate the detection of light atoms with higher sensitivity. However, the conventional ptychography reconstruction may fail when sample-induced multiple scattering is severe.

Therefore, we require an approach to achieve 3D atomic-resolution structural imaging for finite-thickness objects, at reduced electron dose with sufficient sensitivity to resolve light elements. Here, we propose the MultiSlice Electron Tomography (MSET) method to achieve this using tilt series of 4D-STEM datasets. In the MSET algorithm, the nonlinear effect from multiple scattering can be corrected using the inverse multislice approach. Furthermore, unlike previous ptychographic methods, which required two separate reconstruction algorithms (2D ptychographic reconstructions and 3D tomographic reconstruction), our method employs an integrated reconstruction procedure to directly obtain a 3D volume from raw 4D-STEM tilt series (Fig. 1b). This integrated approach enforces a stronger constraint: the diffraction data of all tilt angles should be consistent with a single fixed target 3D phase volume. Therefore, the present method can provide a robust reconstruction fidelity of objects including light elements with much lower electron dose, compared to the methods based on separate 2D ptychographic reconstructions. We numerically demonstrate the superior reconstruction quality of MSET in model systems of CuAu and BaO nanoparticles by testing various experimental parameters including probe defocus, probe step size, electron dose, and tilt angle range. The results consistently show effectively reduced undesirable nonlinearity-based artifacts with better localization precision and detection sensitivity (for example, O atoms next to Ba atoms can be clearly resolved) even at reduced electron dose by a factor of 100 compared to ADF-STEM-based tomography.

## II. METHODS

### A. Atomic structure generation

An atomic structure of a CuAu core-shell nanoparticle of about 4 nm diameter was numerically generated. Initially, a random-shaped 3D volume was generated [9] to place the atoms inside the 3D volume based on the face-centered cubic structure with a lattice constant of 3.78 Å [35]. The total number of atoms was 1602. We also applied a random spatial displacement to each atom, resulting in a root mean square deviation (RMSD) of 20 pm compared to the perfect lattice. To generate the core-shell structure, we applied the alpha-shape algorithm with a shrink factor of 0.9 to separate the core part (576 atoms; assigned as Cu) from the shell part (1026 atoms; assigned as Au) [36]. An atomic structure of a BaO nanoparticle of about 4 nm diameter was created using a similar procedure except for the core-shell separation. The atoms were aligned with a cubic structure with a lattice constant of 5.54 Å [37]. The total number of atoms in this model was 987 (468 for Ba and 501 for O).

### B. Generation of 4D-STEM and ADF-STEM data

To compute the simulated 4D-STEM and ADF-STEM data of the small nanoparticles, we used a multislice method [38]. To solve the scattering problem of fast electrons traveling along the $z$-direction, we considered the paraxial Schrödinger equation [18] described as

$$\frac{\partial \psi(x,y,z)}{\partial z} = \frac{i\lambda}{4\pi}\nabla^2_{xy}\psi(x,y,z) + i\sigma V(x,y,z)\psi(x,y,z), \quad (1)$$

where $\psi$ is the electron wave function, $\lambda$ is the relativistic electron wavelength, $\sigma$ is the relativistic interaction constant, and $V(x,y,z)$ is the 3D electrostatic potential of a specimen. The multislice method is an efficient method to solve Eq. 1 in the STEM configuration [18]. In STEM, a focused electron probe is raster-scanned onto the specimen while measuring the intensity of scattered electrons at each scan point. The wave function of the electron probe after passing through the specimen was obtained as follows. First, the incident electron probe wave function $\psi_{0,\vec{r}_p}$ at each scanned probe position was initialized using the microscope parameters as

$$\psi_{0,\vec{r}_p}(\vec{r}) = \int_0^{q_{max}} \exp(-2\pi i \vec{q} \cdot (\vec{r} - \vec{r}_p)) \exp(-i\chi(q)) \, d\vec{q}, \quad (2)$$

where $\vec{r} = (x,y)$ is a 2D coordinate vector, $\vec{q} = (q_x, q_y)$ is a 2D coordinate vector in Fourier space, and $\vec{r}_p$ is the probe position. $\chi$ is the microscope-dependent aberration function written as

$$\chi(q) = \frac{2\pi}{\lambda}\left(\frac{1}{2}\lambda^2 q^2\, C_1 + \frac{1}{4}\lambda^4 q^4\, C_3 + \frac{1}{6}\lambda^6 q^6\, C_5 + \cdots\right), \quad (3)$$

where $\Delta f = -C_1$ is the defocus and $C_3$ and $C_5$ are third and fifth order spherical aberrations, respectively. Second, the 3D electrostatic potential $V$ was calculated using the Lorentz-Gaussian model based on the parameters tabulated in [18]. The 3D potential was then separated into multiple thin slices of thickness $\Delta z$ along the electron beam direction. Each slice was projected (summed) along the beam direction to form a 2D potential, such that $V_m^{2D}(x,y) = \int_{z_m}^{z_m+\Delta z} V(x,y,z)\, dz$ for all the slices ($m$ is the slice index). In other words, the full 3D potential was approximated into multiple 2D projected potentials. Third, the evolution of the electron wave function was calculated by sequentially solving the free-space propagation part (the first term of the right-hand side of Eq. 1) and the transmission part (the second term of the right-hand side of Eq. 1) for each slice. Accordingly, $\psi_{m+1}$, an electron wave function after transmitting through the $m$-th slice, can be expressed using $\psi_m$ as

$$\psi_{m+1,\vec{r}_p}(\vec{r}) = \mathcal{F}^{-1}[P(\vec{q})\, \mathcal{F}[t_m(\vec{r})\, \psi_{m,\vec{r}_p}(\vec{r})]], \quad (4)$$

where $\mathcal{F}$ and $\mathcal{F}^{-1}$ represent the Fourier transform and its inverse, respectively. The free-space propagation function $P(\vec{q})$ can be written as $P(\vec{q}) = \exp(-i\pi\lambda|\vec{q}|^2 \Delta z)$, and the transmission function $t_m(\vec{r})$ is given as $t_m(\vec{r}) = \exp(i\sigma V_m^{2D}(x,y))$. Note that the thickness $\Delta z$ can be different for each slice in general, but we used a fixed $\Delta z$ for all the slices in our simulations. The exit wave function $\psi_{exit,\vec{r}_p}\, (= \psi_{N_z+1,\vec{r}_p})$, an electron wave function right after transmitting through the specimen, can be obtained by recursively applying Eq. 4 starting from the initial probe wave function to the last slice. $N_z$ is the total number of the slices. The intensity of the exit wave function in Fourier space, $I_{\vec{r}_p}(\vec{q}) = \left|\psi_{\text{exit},\vec{r}_p}(\vec{q})\right|^2$, represents the diffraction data for the probe position $\vec{r}_p$. Integrating the intensity over the ADF detector area gave the ADF image intensity for the probe position (Figs. 2a-c). By applying the calculation for all probe positions, the full 4D-STEM dataset (consisting of 2D diffraction data for probe positions of a 2D raster scan) or an ADF-STEM image was obtained for a given tilt angle. A tomographic tilt series of 4D-STEM or ADF-STEM datasets was generated by repeating this process for different specimen tilt angles.

### C. Reconstruction algorithm (MSET algorithm)

The MSET algorithm solves the inverse problem, which sets out to find the unknown 3D potential $V'$ from a given tilt series of 4D-STEM data (described in Fig. 1b). The algorithm minimizes the error function $\mathcal{E}^2$, which represents the error between the measured 4D-STEM diffraction patterns and the calculated 4D-STEM diffraction patterns from the 3D potential $V'$:

$$V' = \arg\min_{V'} \sum_{\theta} \sum_{\vec{r}_p} \mathcal{E}^2_{\theta,\vec{r}_p}(V') = \arg\min_{V'} \sum_{\theta} \sum_{\vec{r}_p} \left\| \sqrt{\hat{I}_{\theta,\vec{r}_p}(\vec{q};V')} - \sqrt{I_{\theta,\vec{r}_p}(\vec{q})} \right\|_2^2, \qquad (5)$$

where $\|\cdot\|_2$ represents the $L_2$ norm, and $\hat{I}_{\theta,\vec{r}_p}(\vec{q};V')$ and $I_{\theta,\vec{r}_p}(\vec{q})$ are the calculated and measured 4D-STEM diffraction patterns at each probe position $\vec{r}_p$ and each tilt angle $\theta$, respectively.

To obtain the optimal 3D potential $V'$, we implemented a gradient descent method (see the pseudocode in APPENDIX A for more details). Our algorithm consists of three main parts: i) forward propagation, ii) updating $V'$ by backpropagation, and iii) regularization. The forward propagation calculates the 4D-STEM diffraction patterns using the multislice algorithm from $V'$ for each $\vec{r}_p$ and $\theta$. The error function $\mathcal{E}^2$ can be obtained by comparing the estimated and the measured diffraction patterns. By taking the gradient of the error function for each slice of the 3D volume (see APPENDIX B for detailed information about the gradient calculation), the 3D volume can be updated with the gradient descent method in a slice-by-slice manner for every $\vec{r}_p$ and $\theta$ (backpropagation). Then, using *a priori* knowledge that the atomic potential should always be positive and real, a regularization based on positivity was applied by setting all the negative pixels to be zero. These steps repeated until the error $\mathcal{E}^2$ converged. Since the stability of the gradient descent algorithm depends strongly on the step size, we implemented the backtracking line search algorithm [39] in our MSET process to adaptably optimize the step size during the reconstruction. Figure 2d shows an example of the mean error curve for an MSET reconstruction and the error monotonically decreased and converged. Note that by just making the slice thickness $\Delta z$ the same as the specimen thickness (i.e., the volume is represented using a single-slice), our algorithm can be directly used as a conventional single-slice electron tomography (SSET), which is faster than the MSET, but suffers from nonlinearity artifacts as it cannot account for multiple scattering.

**D. Simulation parameters**

In the multislice simulation, the atomic structures of the CuAu and BaO nanoparticles were placed in 3D volumes with a size of 5.12 nm$^3$. We generated the 4D-STEM and ADF-STEM data from the 3D volumes with the following parameters: acceleration voltage of 300 kV, slice thickness of 2 Å, probe step size of 0.4 Å, 0.8 Å, 1.6 Å, 3.2 Å, 6.4 Å, or 12.8 Å (corresponding to 128 × 128, 64 × 64, 32 × 32, 16 × 16, 8 × 8, and 4 × 4 scan positions, respectively), potential sampling pixel size of 0.057 Å (resulting in a diffraction pattern size of 896 × 896 pixels), 21 mrad convergence semi-angle, and 8 frozen phonon configurations. For the ADF-STEM simulations, 40 and 200 mrad were used as the detector inner and outer semi-angles, respectively (Fig. 2c). Poisson noise was added to each diffraction pattern image after adjusting the total number of electrons based on the given electron dose. We generated the tilt series of 20 projections with the angle range of –90° to +90°, –75° to +75°, –50° to +50°, and –25° to +25° depending on the missing wedge parameters.

For ADF-STEM-based 3D reconstruction simulations, we used the GENFIRE and the RESIRE algorithms [40,41] with the following parameters: number of iterations 100, the oversampling ratio of 3, interpolation radius of 0.3 pixels (only for GENFIRE), and discrete Fourier transform interpolation method (only for GENFIRE). For 4D-STEM-based 3D reconstruction simulations, we used both MSET and SSET algorithms with the following parameters: acceleration voltage of 300 kV, probe step size of 0.4 Å, 0.8 Å, 1.6 Å, 3.2 Å, 6.4 Å, or 12.8 Å (corresponding to 128 × 128, 64 × 64, 32 × 32, 16 × 16, 8 × 8, and 4 × 4 scan positions, respectively), potential sampling size of 0.4 Å, the slice thickness of 0.4 Å, the input diffraction image size of 128 × 128 pixels, $10^2$ initial step size of gradient descent. We used the backtracking line search with the parameters of $c = 0$ and $\rho = 0.1$ [39]. The above parameters result in the size of the reconstructed volumes of 128 × 128 × 128 voxels with a voxel size of 0.4 Å.

**E. Atom-tracing and Species classification**

To identify the 3D atomic positions from 3D tomograms reconstructed from 4D-STEM or ADF-STEM data, we applied the atom-tracing method to the 3D tomogram as described previously [6,7]. We classified the traced atoms into three atom types (Au/Cu/non-atom for the CuAu nanoparticle, and Ba/O/non-atom for the BaO nanoparticle) using the species classification method [7]. To prevent misclassification due to the fitting failure for overly sharp potentials, we smoothened the tomograms prior to the species classification using a 3D Gaussian convolution filter with a standard deviation of 0.3 Å. For the BaO nanoparticle, a binary 3D mask was also applied using Otsu's threshold [42] before the species classification to remove the artifacts from undesirable background noise.

**F. Calculation of position error, tracing error, and total classification rate**

To evaluate the performance of the 3D reconstruction algorithms, we defined a position error, a tracing error, and a total classification rate between the traced atomic model and the ground truth atomic model.

i) We calculated the root-mean-square deviation (RMSD) between the positions of traced atoms (excluding non-atoms) and the ground truth atom positions to represent the position error.

ii) We defined false positive traced atoms as the traced atoms (from the reconstruction) not in the ground truth atomic model, and true negative traced atoms as the atoms in the ground truth atomic model but not traced from the reconstruction. Then, the tracing error was calculated as the ratio of the sum of the number of false positive and true negative traced atoms to the total number of atoms in the ground truth atomic model.

iii) We defined the true positive classified atoms as the correctly classified atoms (excluding non-atoms). Finally, the total classification rate was computed as the ratio of the number of true positive classified atoms to the total number of atoms in the ground truth atomic model.

## III. RESULTS

### A. Effect of reconstruction algorithms

We first compared the reconstruction performance of different algorithms and methods: 4D-STEM-based tomography (the MSET and SSET algorithms: Figs. 3a-b) and ADF-STEM-based tomography (the RESIRE and GENFIRE algorithms: Figs. 3c-d) in an infinite-dose, zero-defocus and full tilt range (–90° to +90°) setting with a probe step size of 0.4 Å (corresponds to 128 × 128 probe positions per tilt angle). All results identified 3D individual atomic positions, but MSET yielded a considerably smaller RMSD position error (~6 pm) compared to the other results (~10 pm for SSET, ~12 pm for ADF-STEM results). Remarkably, MSET also provided a zero tracing error (0%) and an almost perfect classification rate (99.9%), which is in sharp contrast to the other linear methods which exhibited about 2% tracing error (about 30 true negative and false positive atoms) and a slightly worse classification rate (~99%). This shows that properly treating nonlinearity via the multislice approach is essential for accurate 3D structural determination.

### B. Effect of probe step sizes and defocus values

Next, we questioned whether overlapping the diffused electron beams of different probe positions can improve the quality of phase retrieval and the reconstruction fidelity in MSET [34,43]. We tested our method at two different defocus values (zero and –20 nm) with various probe step sizes from 0.4 Å (number of scan positions: 128 × 128) to 12.8 Å (number of scan positions: 4 × 4) in an infinite-dose and full tilt range (–90° to +90°) condition. The results in Figs. 4a-f indicates that the reconstruction quality gradually degrades as the number of scan positions decreases. In particular, we found that the proper structural determination fails if the number of scan positions goes below 32 × 32. In such a condition, the RMSD position error becomes larger than 40 pm with over 80% tracing error, and the chemical elements forming the core and shell were misclassified, as can be clearly seen in the intensity histograms and the center slices (Figs. 4d-f). This is because the electron probe is tightly focused at zero defocus condition and the information overlap between the data from the neighboring probe positions is insufficient, resulting in poor phase retrieval [34,43,44]. On the other hand, in the out-of-focus condition (–20 nm), the results from 128 × 128, 64 × 64, and 32 × 32 scan positions show almost identical results in terms of position error (5.6 pm), zero tracing error, and 99.9% classification success rate. Moreover, the 3D reconstructions from the number of scan positions down to 8 × 8 conditions (Figs. 4i-j) could still provide properly identified atomic structures. Note that similar reconstruction was guaranteed regardless of the probe defocus if the data contained a sufficient number of scan positions (~6 pm RMSD position error, zero tracing error, and 99.9% classification rate for 128 × 128 scan positions, Figs. 4a,g). However, due to electron dose limitation for radiation-sensitive materials, it is beneficial to reduce the number of scan positions as much as we can. Our results show that under the appropriately chosen defocus condition, the number of scan positions can be decreased without any penalty, and it can even be further reduced if we sacrifice the reconstruction quality depending on the target precision.

## C. Effect of electron dose levels

Next, we quantified the effect of electron dose on the reconstruction performance. The total electron dose varied from infinity to $2 \times 10^3$ e/Å$^2$ for an entire tilt series in a –20 nm defocus and full tilt range (–90° to +90°) settings, with a probe step size of 0.4 Å (128 × 128 scan positions). In higher electron dose conditions, we observed better RMSD position error, tracing error, and classification rate, as expected (Fig. 5). In the dose conditions below $2 \times 10^6$ e/Å$^2$, the signals in the diffraction images suffered from a strong Poison noise. Nevertheless, the reconstructed tomograms still clarified the atomic structure and accurately classified the atoms unless in the very low dose condition of $2 \times 10^3$ e/Å$^2$ (Figs. 5d-f). Based on the results, we concluded that decent atom identification and classification (zero tracing error and ~95% classification rate) can be achieved for the total electron dose over $2 \times 10^5$ e/Å$^2$.

## D. Effect of the range of tilt angles

In typical electron tomography experiments, a full angle range of –90° to +90° cannot be achieved due to electron beam shadowing and stage tilt limitation. The limitation, also known as the missing wedge problem, restricts the typical angular range for atomic resolution electron tomography to be –75° to 75° or below [4,7,8,10,11]. To see the effect of a limited angular range, we tested the MSET algorithm at four different tilt ranges: –90° to +90°, –75° to +75°, –50° to +50°, and –25° to +25°. The defocus and the probe step size were set to –20 nm and 0.4 Å, respectively. As can be seen in Figs. 6a-c, the reconstructions from the tilt ranges of –90° to +90°, –75° to +75°, and –50° to +50° consistently show properly-determined atomic structures with approximately 6 pm RMSD position error, zero tracing error, and almost perfect classification (~99.9% success rate) (Figs. 6a-c). We interpret that the MSET can robustly retrieve the 3D structural information at experimentally available tilt range conditions. However, as shown in Fig. 6d, the large missing wedge (–25° to +25° angular range) results caused poor axial resolution and undesired artifacts, which resulted in large RMSD position error and tracing error (Fig. 6d). Nevertheless, high-Z elements such as Au were correctly traced and classified even in the presence of such artifacts, suggesting that the MSET could still be useful for very limited tilt range conditions.

## E. Results under typical experimental conditions

Motivated by the successful results of MSET under various imaging conditions, we continued to simulate the 3D reconstructions under a typical ADF-STEM-based atomic electron tomography experiment. In the simulation, we set $2 \times 10^5$ or $2 \times 10^6$ e/Å$^2$ electron dose for an entire tilt series with a tilt range of about –75° to +75° [4,7,8,10]. The defocus and the probe step size were set to –20 nm and 0.4 Å, respectively. We compared the reconstruction results of the CuAu and BaO nanoparticles using MSET, SSET, and RESIRE (Fig. 7). The 3D reconstructions in these experimental conditions suffer from stronger noises and undesired artifacts from the limited electron dose and the missing wedge artifacts compared to the ideal condition results (Fig. 3), but reasonable atomic structures can still be retrieved. As can be seen in Fig. 7, the MSET results are clearly superior to those of SSET or RESIRE. In the case of CuAu nanoparticle reconstructions, although all the algorithms can reasonably detect the positions and chemical species of the Cu and Au atoms, the MSET result (Fig. 7b) shows much better precision in terms of both RMSD and tracing (8.9 pm RMSD position error and 1.3% tracing error) even in the case of $2 \times 10^5$ e/Å$^2$ electron dose per tilt angle, while 10.9 pm RMSD position error and 3.2% tracing error were observed from the SSET reconstruction (Fig. 7d), and the RESIRE result (Fig. 7f) is the worst among those of the three algorithms (17.0 pm RMSD position error and 9.6 % tracing error). Note that the classification success rate in the RESIRE (Fig. 7f) is the best among those of the three algorithms (Figs. 7b,d,f) because ADF-STEM presents stronger contrast for Cu and Au atoms compared to 4D-STEM resulting in the best classification rate.

The difference between different algorithms is more pronounced for the result of the BaO nanoparticle. As mentioned earlier, a typical ADF-STEM-based tomography cannot easily identify low-Z elements (such as O atom) especially when there are heavier elements (such as Ba atom) mixed within the specimen. Figures 7g-l clearly show this: a majority of oxygen atoms are missing in the SSET and RESIRE reconstructions (even at $2 \times 10^6$ e/Å$^2$ electron dose), and even in the MSET results if the electron does is not enough ($2 \times 10^5$ e/Å$^2$ electron dose case: Fig. 7h). Only when sufficient electrons are used for data acquisition ($2 \times 10^6$ e/Å$^2$ electron dose), the oxygen atoms can be reliably detected using the MSET algorithm (Fig. 7g). Our result suggests that strong nonlinearity can negatively affect the intensity of light elements, making it similar to the background noise and artifacts if a linear projection assumption is used in the reconstruction (SSET and RESIRE), which critically hinders proper classification between the artifacts and low-Z elements. Therefore, it is necessary to properly treat the nonlinear effect to accurately determine the 3D atomic structure using electron tomography (especially for low-Z elements), which is possible using our MSET algorithm.

We finally leveraged the robust reconstruction performance of MSET by further reducing the number of scan positions. We reduced the number of scan positions from $128 \times 128$ to $32 \times 32$, which economized the electron dose 16 times (Fig. 8). The result of the CuAu shows that MSET successfully maintained the reconstruction fidelity (Figs. 8a-b). At the lowest dose condition (total electron dose of the tilt series: $1.25 \times 10^4$ e/Å$^2$), the reconstructed tomogram using MSET provided the RMSD position error of 10.1 pm, tracing error of 0.7%, and total classification rate of 97.3%. This performance is better than the ADF-STEM result in Fig. 7e which used approximately 100 times more electron dose (total electron dose of the tilt series: $2 \times 10^6$ e/Å$^2$). Also, in the case of the BaO nanoparticle (Figs. 8c-d), the oxygen atoms were well-detected with similar sensitivity compared to the larger dose case ($128 \times 128$ scan positions: Figs. 7g-h). In short, the 4D-STEM-based MSET algorithm allows us to reduce the electron dose significantly (by a factor of 100) while maintaining superior detection capability in terms of RMSD, tracing error, and classification rate.

## F. Summary of the simulation results

Figure 9 compares the position error (RMSD), tracing error, and classification success rate of the simulation results from the CuAu nanoparticle given in Figs. 4-8. Again, MSET outperforms the linearity-based algorithms in typical experimental conditions (Figs. 9j-l), and it is also clear from the figure that the quality of retrieved atomic structure strongly depends on the parameters such as electron dose, probe defocus value, number of scan positions, and angular range, as already discussed above. They should be properly chosen under the given experimental limitations (available angular range, maximum electron dose for given specimen, etc.) to obtain the best reconstruction results.

## IV. CONCLUSIONS

We developed MSET, a new 4D-STEM-based tomography method using a multislice algorithm to determine the 3D atomic structure of various nanomaterials, with improved precision and accuracy at a lower electron dose. Our method can avoid undesirable nonlinear effects which cannot be resolved by conventional ADF-STEM-based tomography or single-slice-based tomography. Our simulations suggest that choosing a proper defocus condition and number of scan positions can drastically reduce electron dose without loss of reconstruction quality. In the simulations with a low electron dose and a typical missing wedge condition, we demonstrated that MSET can determine the 3D structure of the CuAu core-shell nanoparticle with 10.1 pm precision and elemental selectivity determined with 97.3% accuracy. Our simulations of a BaO nanoparticle showed that the 3D coordinates of light elements (O atoms in this case) can be precisely detected. We expect that MSET will provide a new pathway to determine unknown 3D atomic structures of various nanomaterials including radiation-sensitive samples and materials which contain light elements. We will publicly release the source codes of MSET [45].

## ACKNOWLEDGEMENTS

This research was mainly supported by Samsung Science and Technology Foundation (SSTF-BA2201-05). M.L. and Y.K.P. were supported by National Research Foundation of Korea (2015R1A3A2066550). Work at the Molecular Foundry was supported by the Office of Science, Office of Basic Energy Sciences, of the U.S. Department of Energy under Contract No. DE-AC02-05CH11231. C.O. acknowledges support from a US DOE Early Career Research Award.

# APPENDIX A. PSEUDO CODE FOR MSET ALGORITHM

**PSEUDO CODE: MSET ALGORITHM TO RECONSTRUCT 3D POTENTIAL**

**input**: A tilt series of measured full diffraction pattern images (4D-STEM dataset) $\{I_{\theta,\vec{r}_p}\}$, Tilt angle set $\{\theta\}$, Probe position set $\{\vec{r}_p\}$, Step size of the gradient descent method α, and Number of iterations $N_{iter}$

**output**: 3D potential $V'(x, y, z)$

1. **Initialization:** $V'_{(1)}(x, y, z) \leftarrow 0$

   /* main reconstruction */

2. $P(q_x, q_y) \leftarrow$ free-space propagation with thickness $\Delta z$ (Eq. 4)
3. **for** i in $\{1$ to $N_{iter}\}$ **do**
4.     **for** θ in tilt angle set $\{\theta\}$ **do**
5.         $V'_{rot}(x, y, z) \leftarrow R_\theta V'_{(i)}(x, y, z)$     // $R_\theta$: rotation matrix by θ tilt angle
6.         **for** $\vec{r}_p$ in probe position set $\{\vec{r}_p\}$ **do**

               /* forward propagation */

7.             $\psi_{1,\vec{r}_p}(x, y) \leftarrow$ initial probe beam function at $\vec{r}_p$ (Eq. 2)
8.             $\{V_m^{2D}(x, y)\} \leftarrow$ projected potential set from $V'_{rot}(x, y, z)$ (Eq. 4)
9.             $\{t_m(x, y)\} \leftarrow$ transmission function set from $\{V_m^{2D}(x, y)\}$ (Eq. 4)
10.            **for** m in $\{1$ to $N_z\}$ **do**     // $N_z$ represents the last slice
11.                 $\psi_{m+1,\vec{r}_p}(x, y) \leftarrow \mathcal{F}^{-1}[P(q_x, q_y)\, \mathcal{F}[t_m(x, y)\psi_{m,\vec{r}_p}(x, y)]]$     // multislice solution (Eq. 4)
12.            **end for**
13.             $\hat{I}_{\theta,\vec{r}_p}(q_x, q_y) \leftarrow |\mathcal{F}(\psi_{exit,\vec{r}_p}(x, y))|^2$    $(\psi_{exit,\vec{r}_p}(x, y) = \psi_{N_z+1,\vec{r}_p}(x, y))$     // calculate diffraction images

               /* backpropagation */

14.             $\psi_{N_z+1,\vec{r}_p}^{back}(q_x, q_y) \leftarrow \mathcal{F}(\psi_{exit,\vec{r}_p}(x, y))(1 - \sqrt{I_{\theta,\vec{r}_p}(q_x, q_y)} / \sqrt{\hat{I}_{\theta,\vec{r}_p}(q_x, q_y)})$
15.            **for** m in $\{N_z$ to $1\}$ **do**
16.                 $\psi_{m,\vec{r}_p}^{back}(x, y) \leftarrow \mathcal{F}^{-1}[P_m^*(q_x, q_y)\, \psi_{m+1,\vec{r}_p}^{back}(q_x, q_y)]$
17.                 $\nabla_{V'}\mathcal{E}^2(x, y, m) \leftarrow i\, t_m^*(x, y)\psi_m^*(x, y)\psi_{m,\vec{r}_p}^{back}(x, y)$     // compute gradient of m-th slice
18.                 $\psi_{m,\vec{r}_p}^{back}(q_x, q_y) \leftarrow \mathcal{F}[t_m^*(x, y)\, \psi_{m,\vec{r}_p}^{back}(x, y)]$
19.            **end for**

               /* update 3D potential */

20.             $V'_{rot}(x, y, z) \leftarrow V'_{rot}(x, y, z) - \alpha\, \nabla_{V'}\mathcal{E}^2(x, y, z)$
21.         **end for**
22.         $V'_{(i)}(x, y, z) \leftarrow R_\theta^{-1} V'_{rot}(x, y, z)$     // $R_\theta^{-1}$: inverse rotation matrix by θ tilt angle
23.         $V'_{(i)}(x, y, z) \leftarrow 0$ if $V'_{(i)}(x, y, z) < 0$     // positivity constraint
24.     **end for**
25. **end for**
26. $V'(x, y, z) \leftarrow V'_{(N_{iter})}(x, y, z)$

# APPENDIX B. GRADIENT CALCULATION FOR MSET ALGORITHM

We derived the gradient in the MSET algorithm following the vectorized notation and idea from Ren *et al.* [22]. In the vectorized notation, all 2D functions (such as wave function and transmission function) sampled in $N_x \times N_y$ pixels were expressed as 1D column vectors with size $N_x N_y \times 1$, and the linear operators such as Fourier transform were represented as $N_x N_y \times N_x N_y$ square matrices. We use the following symbols and conventions: $\dagger$ is the Hermitian operator, $*$ is the complex conjugate, and diag($\cdot$) is a diagonalization operator which converts a 1D vector into a 2D square matrix by putting the components of the vector into diagonal components in the square matrix, and Re($\cdot$) is an operator which takes only the real part of the complex variables. In addition, the wave function $\boldsymbol{\psi}$, the transmission function $\boldsymbol{t}$, 3D potential $V$ are expressed in real space and the diffraction pattern $\boldsymbol{I}$ is expressed in Fourier space. The error function $\mathcal{E}^2_{\theta, \vec{r}_p}$ for each probe position $\vec{r}_p$ and each tilt angle $\theta$ in Eq. 5 can be expressed in a vectorized form as

$$\mathcal{E}^2_{\theta, \vec{r}_p}(V) = \boldsymbol{\mathcal{E}}^\dagger_{\theta, \vec{r}_p}(V) \boldsymbol{\mathcal{E}}_{\theta, \vec{r}_p}(V), \tag{B1}$$

where $\boldsymbol{\mathcal{E}}_{\theta, \vec{r}_p} \equiv \sqrt{\hat{\boldsymbol{I}}_{\theta, \vec{r}_p}} - \sqrt{\boldsymbol{I}_{\theta, \vec{r}_p}}$ and $\hat{\boldsymbol{I}}_{\theta, \vec{r}_p} = \left|\mathcal{F}(\boldsymbol{\psi}_{\text{exit}, \theta, \vec{r}_p})\right|^2$ (see Eq. 4). $\hat{\boldsymbol{I}}_{\theta, \vec{r}_p}$ and $\boldsymbol{I}_{\theta, \vec{r}_p}$ are the calculated and measured diffraction pattern images at each probe position $\vec{r}_p$ and each tilt angle $\theta$, respectively.

We calculated the derivate of $\mathcal{E}^2_{\theta, \vec{r}_p}(V)$ with respect to the $m$-th slice $\boldsymbol{V}_m$ of 3D potential $V$ as

$$\nabla_{\boldsymbol{V}_m} \mathcal{E}^2_{\theta, \vec{r}_p} \equiv \left[\frac{\partial(\boldsymbol{\mathcal{E}}^\dagger_{\theta, \vec{r}_p} \boldsymbol{\mathcal{E}}_{\theta, \vec{r}_p})}{\partial \boldsymbol{V}_m}\right]^\dagger = \left[\frac{\partial(\boldsymbol{\mathcal{E}}^\dagger_{\theta, \vec{r}_p} \boldsymbol{\mathcal{E}}_{\theta, \vec{r}_p})}{\partial \boldsymbol{\mathcal{E}}_{\theta, \vec{r}_p}} \frac{\partial \boldsymbol{\mathcal{E}}_{\theta, \vec{r}_p}}{\partial \boldsymbol{V}_m}\right]^\dagger. \tag{B2}$$

The first factor in Eq. B2 can be easily calculated as

$$\frac{\partial(\boldsymbol{\mathcal{E}}^\dagger_{\theta, \vec{r}_p} \boldsymbol{\mathcal{E}}_{\theta, \vec{r}_p})}{\partial \boldsymbol{\mathcal{E}}_{\theta, \vec{r}_p}} = 2\boldsymbol{\mathcal{E}}^\dagger_{\theta, \vec{r}_p}. \tag{B3}$$

Using the chain rule, the second factor in Eq. B2 can also be calculated as

$$\begin{aligned}\frac{\partial \boldsymbol{\mathcal{E}}_{\theta, \vec{r}_p}}{\partial \boldsymbol{V}_m} &= \frac{\partial\left[\sqrt{\left|\mathcal{F}(\boldsymbol{\psi}_{\text{exit},\theta,\vec{r}_p})\right|^2}\right]}{\partial\left[\left|\mathcal{F}(\boldsymbol{\psi}_{\text{exit},\theta,\vec{r}_p})\right|^2\right]} \left(\frac{\partial[\text{diag}(\mathcal{F}(\boldsymbol{\psi}_{\text{exit},\theta,\vec{r}_p})^*)\mathcal{F}(\boldsymbol{\psi}_{\text{exit},\theta,\vec{r}_p})]}{\partial[\mathcal{F}(\boldsymbol{\psi}_{\text{exit},\theta,\vec{r}_p})]} \frac{\partial[\mathcal{F}(\boldsymbol{\psi}_{\text{exit},\theta,\vec{r}_p})]}{\partial \boldsymbol{V}_m} + \frac{\partial[\text{diag}(\mathcal{F}(\boldsymbol{\psi}_{\text{exit},\theta,\vec{r}_p}))\mathcal{F}(\boldsymbol{\psi}_{\text{exit},\theta,\vec{r}_p})^*]}{\partial[\mathcal{F}(\boldsymbol{\psi}_{\text{exit},\theta,\vec{r}_p})^*]} \frac{\partial[\mathcal{F}(\boldsymbol{\psi}_{\text{exit},\theta,\vec{r}_p})^*]}{\partial \boldsymbol{V}_m}\right) \\ &= \text{diag}\left(\frac{1}{\sqrt{\left|\mathcal{F}(\boldsymbol{\psi}_{\text{exit},\theta,\vec{r}_p})\right|^2}}\right) \text{Re}\left(\frac{\partial[\text{diag}(\mathcal{F}(\boldsymbol{\psi}_{\text{exit},\theta,\vec{r}_p})^*)\mathcal{F}(\boldsymbol{\psi}_{\text{exit},\theta,\vec{r}_p})]}{\partial[\mathcal{F}(\boldsymbol{\psi}_{\text{exit},\theta,\vec{r}_p})]} \frac{\partial[\mathcal{F}(\boldsymbol{\psi}_{\text{exit},\theta,\vec{r}_p})]}{\partial \boldsymbol{V}_m}\right) \\ &= \text{diag}\left(\frac{1}{\sqrt{\left|\mathcal{F}(\boldsymbol{\psi}_{\text{exit},\theta,\vec{r}_p})\right|^2}}\right) \text{Re}\left(\frac{\partial[\text{diag}(\mathcal{F}(\boldsymbol{\psi}_{\text{exit},\theta,\vec{r}_p})^*)\mathcal{F}(\boldsymbol{\psi}_{\text{exit},\theta,\vec{r}_p})]}{\partial[\mathcal{F}(\boldsymbol{\psi}_{\text{exit},\theta,\vec{r}_p})]} \frac{\partial[\mathcal{F}(\boldsymbol{\psi}_{\text{exit},\theta,\vec{r}_p})]}{\partial \boldsymbol{\psi}_{N_z,\theta,\vec{r}_p}} \frac{\partial \boldsymbol{\psi}_{N_z,\theta,\vec{r}_p}}{\partial \boldsymbol{\psi}_{N_z-1,\theta,\vec{r}_p}} \cdots \frac{\partial \boldsymbol{\psi}_{m+1,\theta,\vec{r}_p}}{\partial \boldsymbol{t}_m} \frac{\partial \boldsymbol{t}_m}{\partial \boldsymbol{V}_m}\right) \\ &= \text{diag}\left(\frac{1}{\sqrt{\left|\mathcal{F}(\boldsymbol{\psi}_{\text{exit},\theta,\vec{r}_p})\right|^2}}\right) \text{Re}\left(\left[\text{diag}(\mathcal{F}(\boldsymbol{\psi}_{\text{exit},\theta,\vec{r}_p})^*)\right][\mathcal{F}\mathcal{F}^{-1}P\mathcal{F}\text{diag}(\boldsymbol{t}_{N_z})][\mathcal{F}^{-1}P\mathcal{F}\text{diag}(\boldsymbol{t}_{N_z-1})]\cdots[\mathcal{F}^{-1}P\mathcal{F}\text{diag}(\boldsymbol{\psi}_{m,\theta,\vec{r}_p})][i\sigma\Delta z\text{diag}(\boldsymbol{t}_m)]\right),\end{aligned} \tag{B4}$$

where $\boldsymbol{\psi}_{m+1,\theta,\vec{r}_p} = \mathcal{F}^{-1}P\mathcal{F}\,\text{diag}(\boldsymbol{t}_{N_z-1})\,\boldsymbol{\psi}_{m,\theta,\vec{r}_p}$ and $\boldsymbol{t}_m = \exp(i\sigma\Delta z \boldsymbol{V}_m)$ (see Eq. 4).

Using the results of Eq. B3 and Eq. B4, the gradient $\nabla_{V_m}\mathcal{E}^2_{\theta,\vec{r}_p}$ can be expressed as

$$\nabla_{V_m}\mathcal{E}^2_{\theta,\vec{r}_p} = Re\left(-2i\sigma\Delta z \mathrm{diag}(t^*_m)\mathrm{diag}(\psi^*_m)\mathcal{F}^{-1}P^*\mathcal{F}\cdots \mathrm{diag}(t^*_{N_z-1})\mathcal{F}^{-1}P^*\mathcal{F}\mathrm{diag}(t^*_{N_z})\mathcal{F}^{-1}P^*\mathcal{F}\mathcal{F}^{-1}\right.$$

$$\left.\left(\mathcal{F}\left(\psi_{\mathrm{exit},\theta,\vec{r}_p}\right) - \mathrm{diag}\left(\sqrt{I_{\theta,\vec{r}_p}}\right)\mathrm{diag}\left(\left|\mathcal{F}\left(\psi_{\mathrm{exit},\theta,\vec{r}_p}\right)\right|^{-1}\right)\mathcal{F}\left(\psi_{\mathrm{exit},\theta,\vec{r}_p}\right)\right)\right). \tag{B5}$$

In our code [45], the constant value $2\sigma\Delta z$ in Eq. B5 is set to unity because we can adjust the magnitude of the gradient using the step size. We calculated the gradient for all slices (from 1 to the $N_z$ slice), which means that the gradient for the full 3D potential $V'$ ($N_x \times N_y \times N_z$ voxels) is determined. After calculating the gradient, the unknown 3D potential $V'$ can be updated using the gradient calculated for the given probe position and tilt angle as

$$V'_{(i+1)} = V'_{(i)} - \alpha_{(i)}\nabla_{V'}\mathcal{E}^2_{\theta,\vec{r}_p}, \tag{B6}$$

where $i$ is the iteration number, and $\alpha_{(i)}$ is the step size at the $i$-th iteration.

# Figures

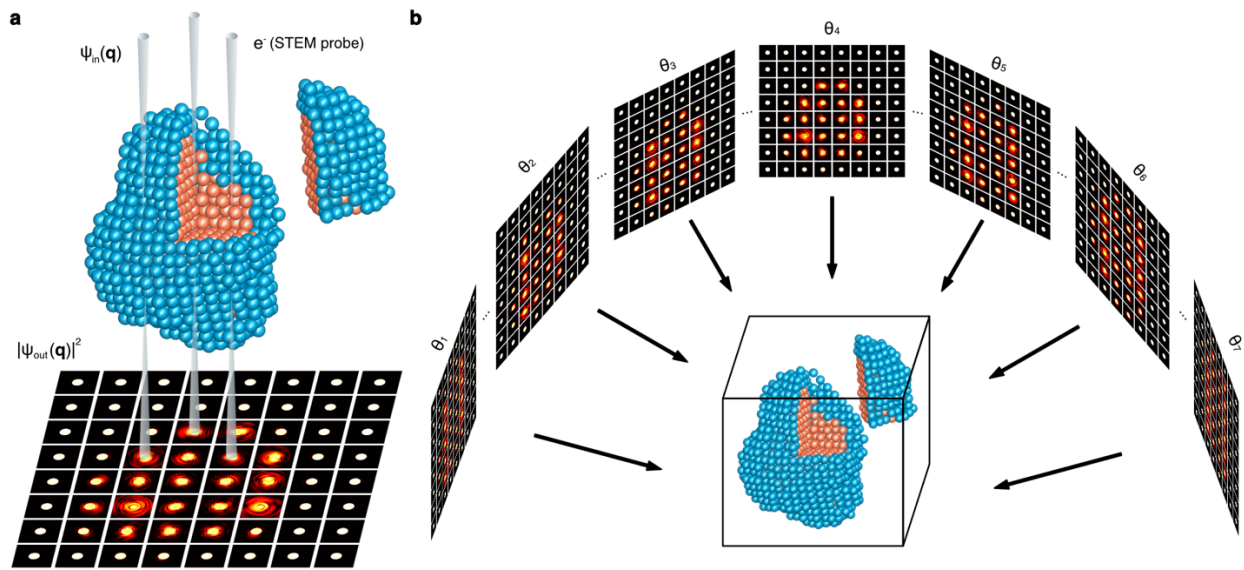

FIG. 1. a. A schematic figure showing the 4D-STEM data acquisition. During the 2D raster of the electron probe, 2D diffraction patterns are recorded for each probe position. b. Schematic of MSET method. From a tilt series of 4D-STEM dataset taken at multiple tilt angles, a 3D atomic potential can be directly retrieved via the multislice approach.

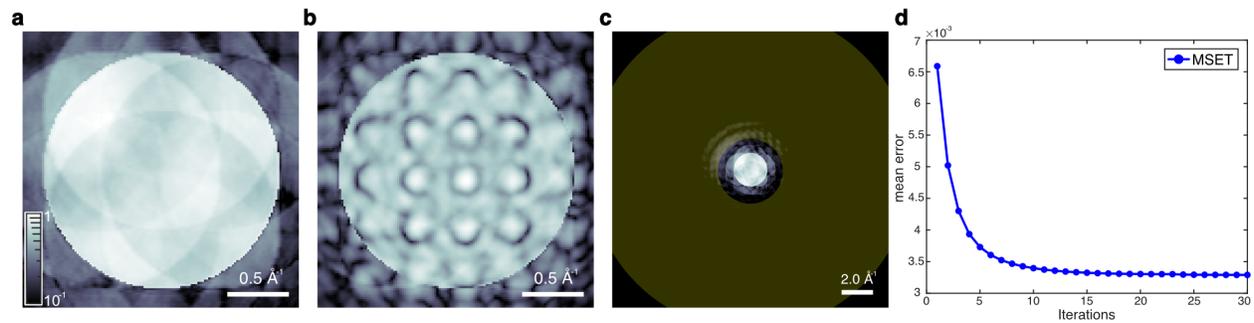

FIG. 2. a-b. Representative 4D-STEM diffraction pattern images when the probe is at the center of the specimen volume at zero defocus condition (a) and –20 nm defocus value condition (b). c. A larger field-of-view diffraction image of (a), which was used for the ADF-STEM simulations. The yellow area in (c) shows the virtual ADF detector area. The diffraction pattern images are plotted on a logarithmic scale. d. The error curve of a representative MSET reconstruction. The mean error is calculated by averaging the error $\mathcal{E}^2$ (defined in Eq. 5) over all tilt angles and probe positions.

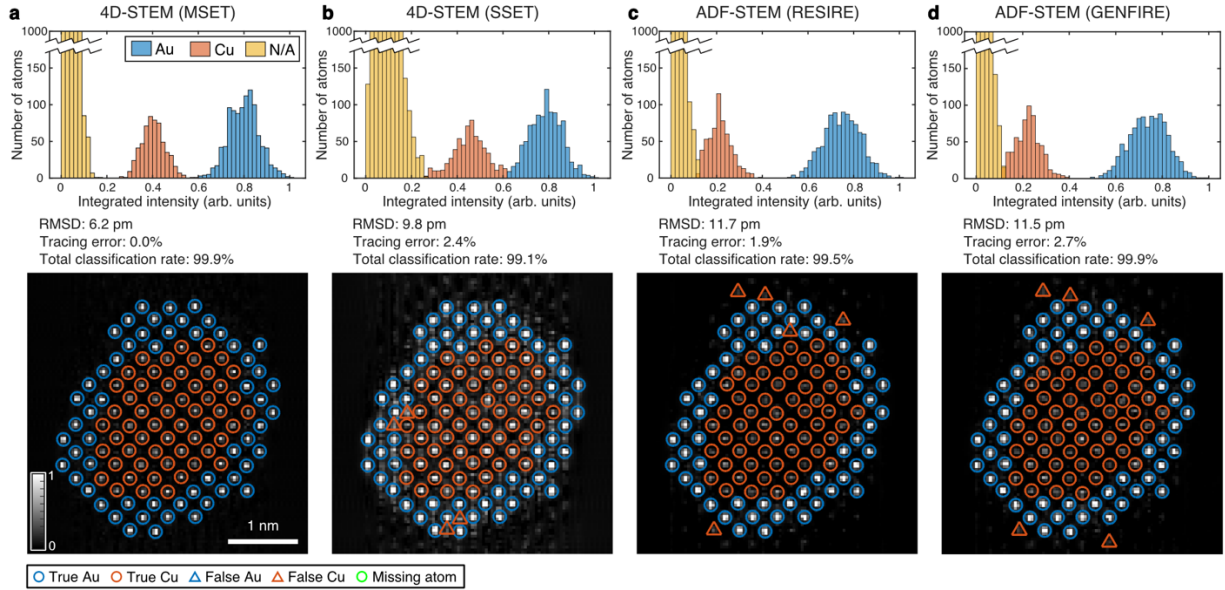

FIG. 3. a-d. 3D reconstruction results from 4D-STEM-based tomography (the MSET and SSET algorithms) and ADF-STEM-based tomography (the RESIRE and GENFIRE algorithms). The intensity histograms (which also show classification results) of traced atoms are shown in the top row, and 2 Å-thick central slices of the 3D reconstructed volumes are shown in the bottom row where the traced atom positions are overlaid. The panels (a-d) represent the results from MSET algorithm (a), SSET algorithm (b), RESIRE algorithm (c), and the GENFIRE algorithm (d). The true and false atoms represent correctly and incorrectly classified atoms, respectively. The missing atom means an atom in the ground truth but not traced or misclassified as a non-atom.

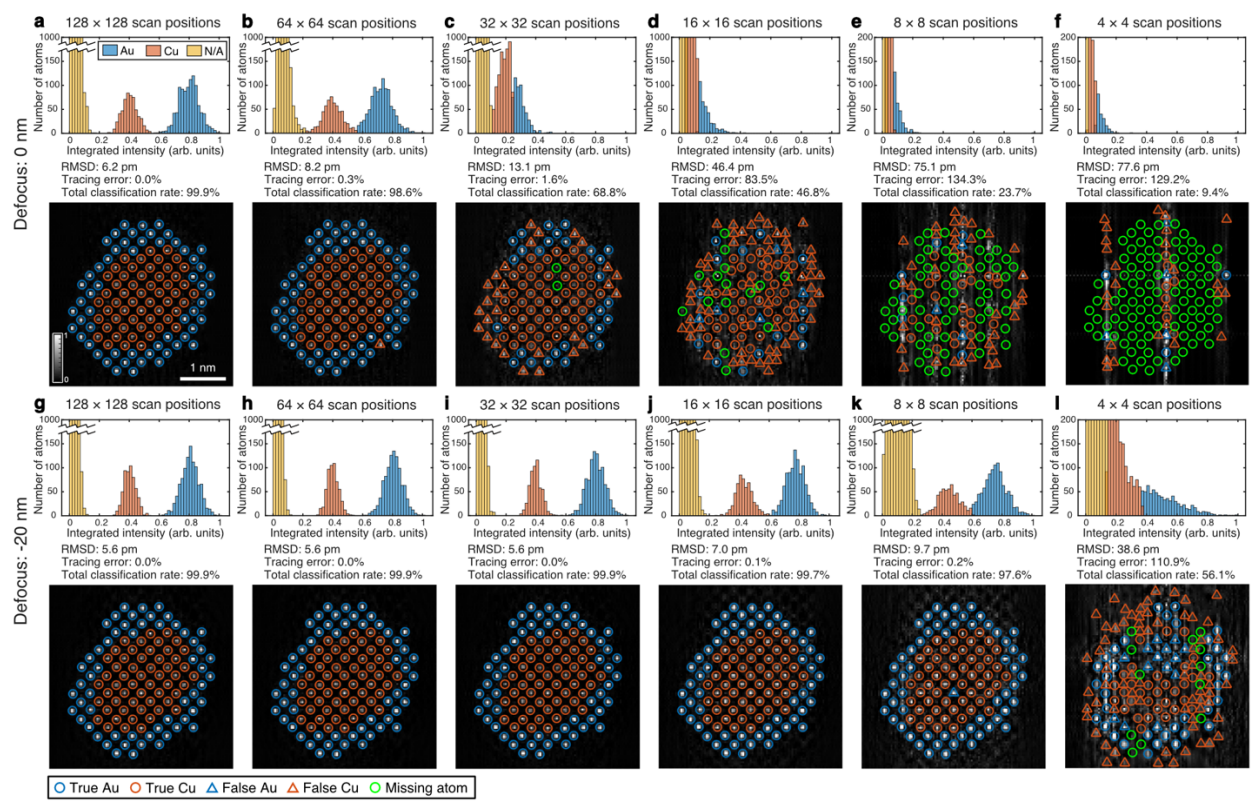

FIG. 4. a-l. 3D reconstruction results from the MSET algorithm from the data of different probe step sizes and defocus values. The intensity histograms (which also show classification results) of traced atoms are shown in the top part of each panel, and 2 Å-thick central slices of the 3D reconstructed volume are shown in the bottom part of each panel, where the traced atom positions are overlaid. The panels (a)-(f) represent the zero-defocus results with the probe step size of 0.4 Å (number of scan positions: 128 × 128) (a), 0.8 Å (number of scan positions: 64 × 64) (b), 1.6 Å (number of scan positions: 32 × 32) (c), 3.2 Å (number of scan positions: 16 × 16) (d), 6.4 Å (number of scan positions: 8 × 8) (e), and 12.8 Å (number of scan positions: 4 × 4) (f). The panels (g)-(l) represent the –20 nm defocus results with the probe step size of 0.4 Å (number of scan positions: 128 × 128) (g), 0.8 Å (number of scan positions: 64 × 64) (h), 1.6 Å (number of scan positions: 32 × 32) (i), 3.2 Å (number of scan positions: 16 × 16) (j), 6.4 Å (number of scan positions: 8 × 8) (k), and 12.8 Å (number of scan positions: 4 × 4) (l). We used the same definition of the true, false, and missing atoms as in Fig. 3.

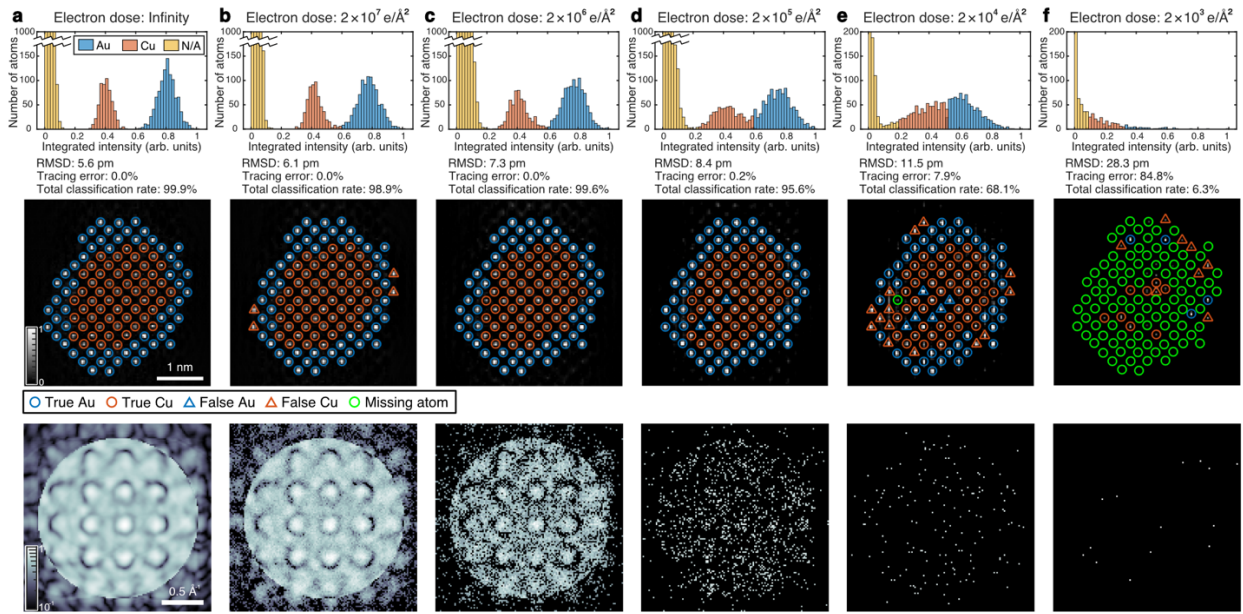

FIG. 5. a-f. 3D reconstruction results from the MSET algorithm from the data of different electron dose levels. The intensity histograms (which also show classification results) of traced atoms are shown in the top row, 2 Å-thick central slices of the 3D reconstructed volume are shown in the middle row where the traced atom positions are overlaid, and the diffraction pattern images at center probe position are shown in the bottom row. The panels (a-f) represent the results with electron dose levels of infinity (a), $2 \times 10^7$ e/Å$^2$ (b), $2 \times 10^6$ e/Å$^2$ (c), $2 \times 10^5$ e/Å$^2$ (d), $2 \times 10^4$ e/Å$^2$ (e), and $2 \times 10^3$ e/Å$^2$ (f) for the entire tilt series. The diffraction pattern images are plotted on a logarithmic scale. We used the same definition of the true, false, and missing atoms as in Fig. 3.

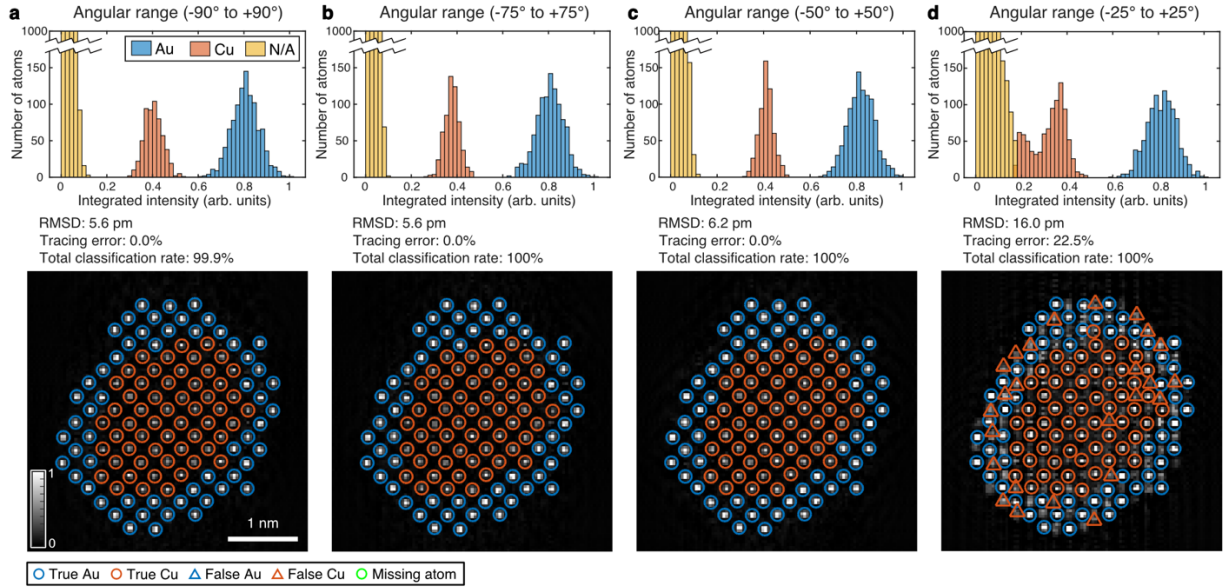

FIG. 6. a-d. 3D reconstruction results from the MSET algorithm from the data of different tilt angle ranges. The intensity histograms (which also show classification results) of traced atoms are shown in the top row, and 2 Å-thick central slices of the 3D reconstructed volume are shown in the bottom row where the traced atom positions are overlaid. The panels (a-d) represent the results with tilt ranges of –90° to +90° (a), –75° to +75° (b), –50° to +50° (c), and –25° to +25° (d). We used the same definition of the true, false, and missing atoms as in Fig. 3.

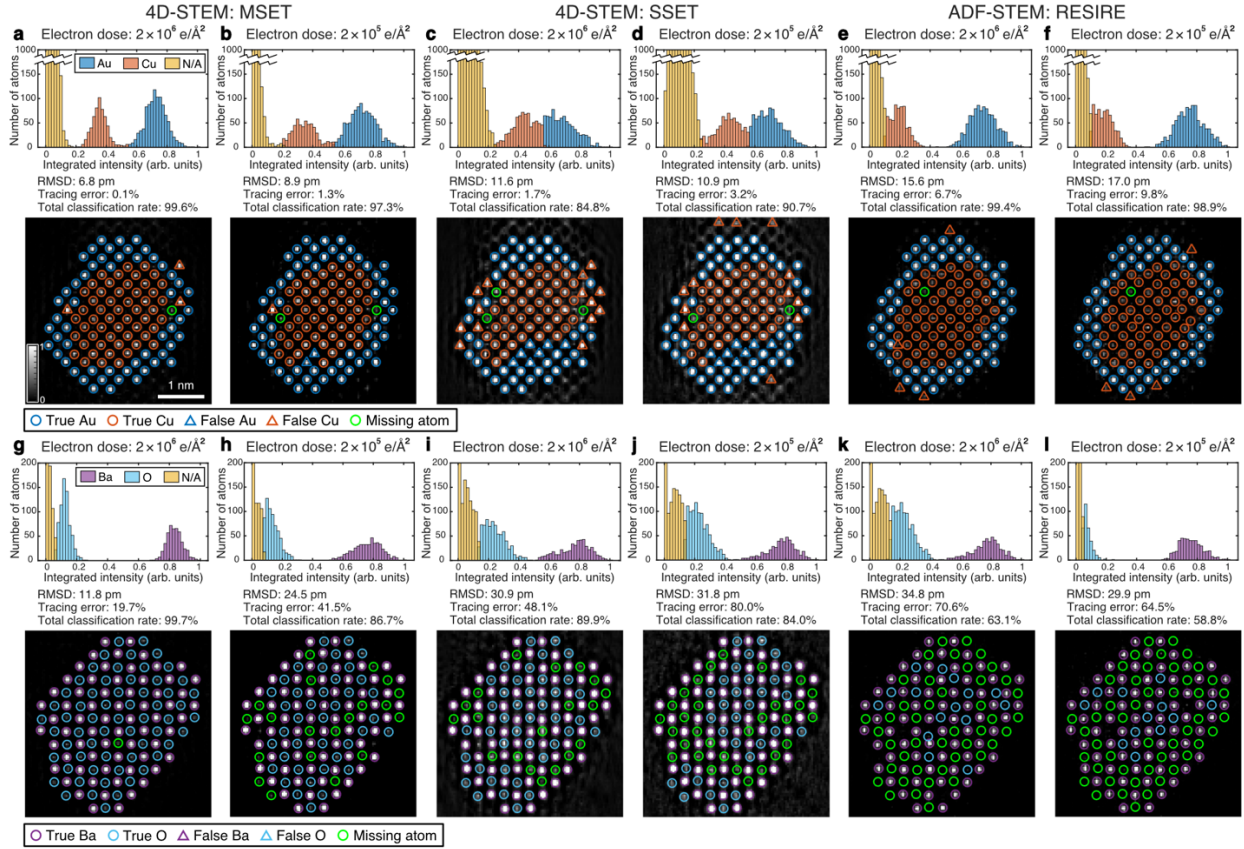

FIG. 7. a-l. 3D reconstruction results of the CuAu and BaO nanoparticles using the different algorithms from the data of different electron dose levels. The probe step size of 0.4 Å (number of scan positions: 128 × 128) and the tilt range of –75° to +75° were used. The intensity histograms (which also show classification results) of traced atoms are shown in the top part of each panel, and 2 Å-thick central slices of the 3D reconstructed volume are shown in the bottom part of each panel where the traced atom positions are overlaid. The panels (a-f) represent the results of the CuAu nanoparticle from [1) the MSET algorithm using the data of $2 \times 10^6$ e/Å$^2$ (a) and $2 \times 10^5$ e/Å$^2$ (b) electron doses for the entire tilt series, 2) the SSET algorithm using data of $2 \times 10^6$ e/Å$^2$ (c) and $2 \times 10^5$ e/Å$^2$ (d) electron doses for the entire tilt series, 3) the RESIRE algorithm using data of $2 \times 10^6$ e/Å$^2$ (e) and $2 \times 10^5$ e/Å$^2$ (f) electron doses for the entire tilt series]. The panels (g-l) represent the results of the BaO nanoparticle from [1) the MSET algorithm using data of $2 \times 10^6$ e/Å$^2$ (g) and $2 \times 10^5$ e/Å$^2$ (h) electron doses for the entire tilt series, 2) the SSET algorithm using data of $2 \times 10^6$ e/Å$^2$ (i) and $2 \times 10^5$ e/Å$^2$ (j) electron doses for the entire tilt series, 3) the RESIRE algorithm using data of $2 \times 10^6$ e/Å$^2$ (k) and $2 \times 10^5$ e/Å$^2$ (l) electron doses for the entire tilt series]. We used the same definition of the true, false, and missing atoms as in Fig. 3.

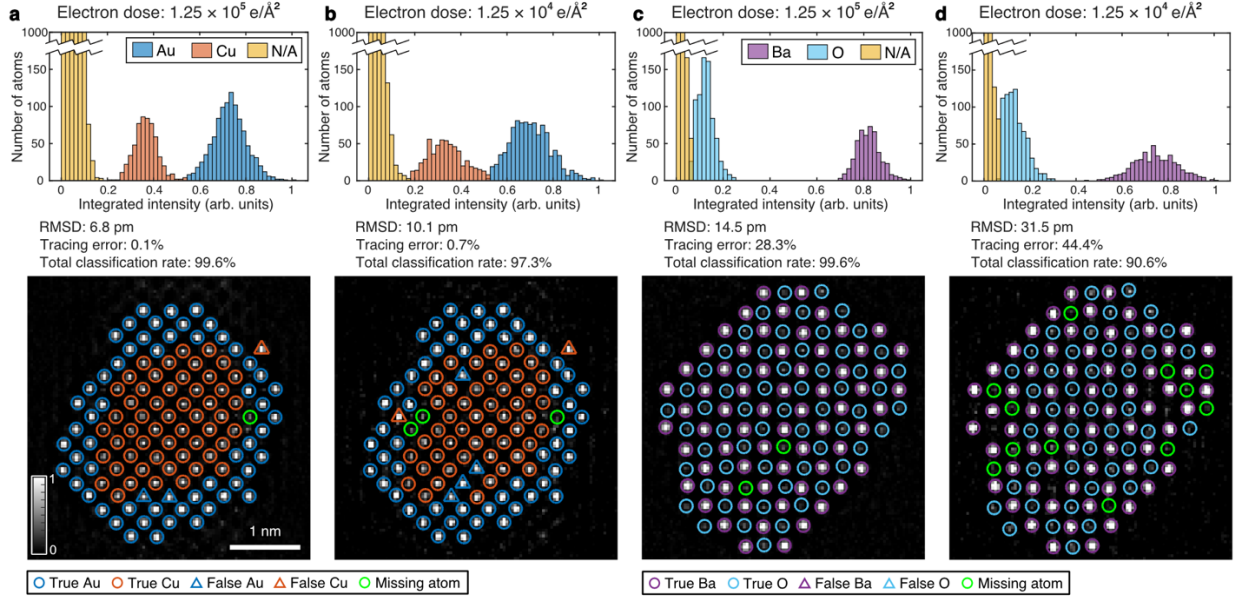

FIG. 8. a-d. 3D reconstruction results of the CuAu and BaO nanoparticles from the MSET algorithm using the data of different electron dose levels. Probe step size of 1.6 Å (number of scan positions: 32 × 32) and the tilt range of –75° to +75° were used. The intensity histograms (which also show classification results) of traced atoms are shown in the top row, and 2 Å-thick central slices of the 3D reconstructed volume are shown in the bottom row where the traced atom positions are overlaid. The panels (a-b) represent the results of the CuAu nanoparticle using the data of electron dose levels of $1.25 \times 10^5$ e/Å$^2$ (a) and $1.25 \times 10^4$ e/Å$^2$ (b) for the entire tilt series. The panels (c-d) represent the results of the BaO nanoparticle using the data of electron dose levels of $1.25 \times 10^5$ e/Å$^2$ (c) and $1.25 \times 10^4$ e/Å$^2$ (d) for the entire tilt series. We used the same definition of the true, false, and missing atoms as in Fig. 3.

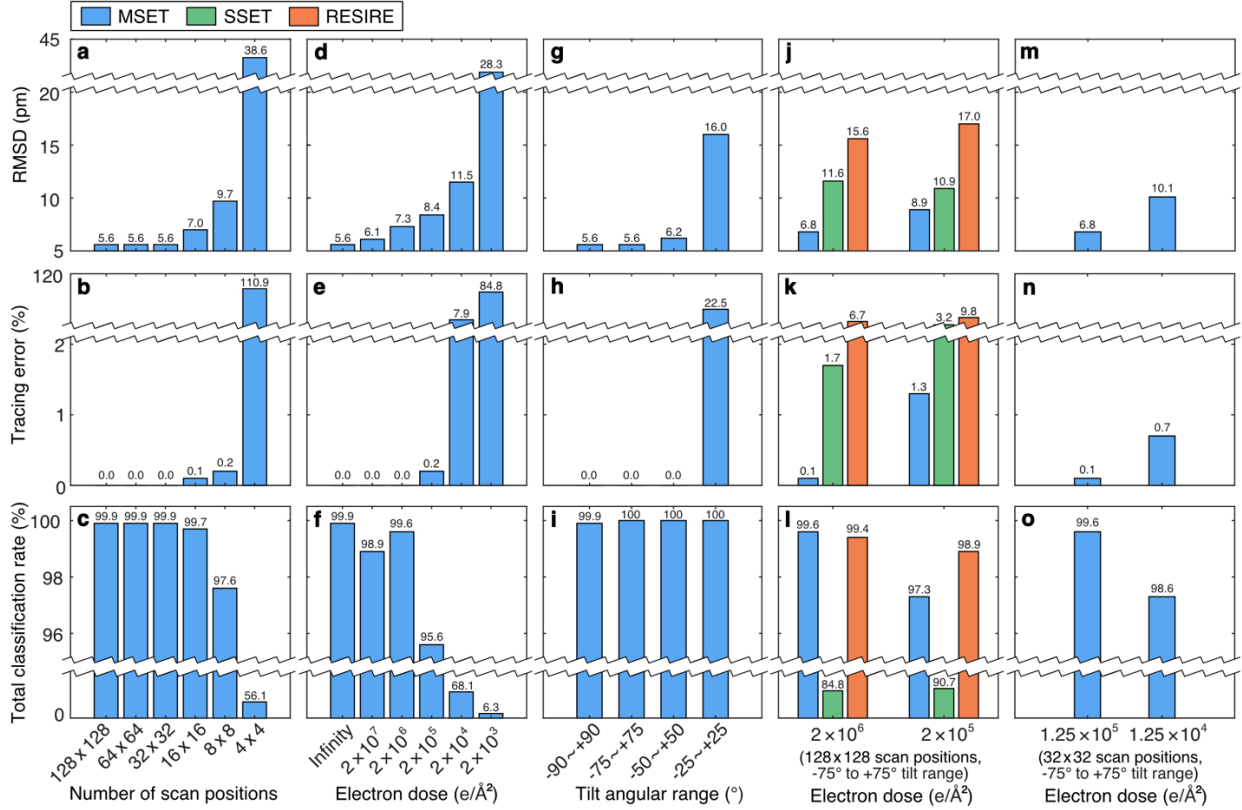

FIG. 9. a-o. Summary of the reconstruction results from the CuAu nanoparticle simulations, showing RMSD position error, tracing error, and total classification error. (a-c) The effect of probe step size (number of scan positions, see Fig 4) for MSET. The effect of electron dose levels for MSET (Fig. 5), (g-i) The effect of the range of tilt angles for MSET (Fig. 6). (j-l) The effect of different reconstruction algorithms (MSET, SSET, and RESIRE) under typical experimental conditions with 128 × 128 scan positions (Fig. 7). (m-o) MSET results under typical experimental conditions with 32 × 32 scan positions (Fig. 8)